\documentclass[twocolumn,amsmath,amssymb,prl,preprintnumbers]{revtex4}
\usepackage{graphicx}
\usepackage{dcolumn}
\usepackage{bm}
\usepackage{amsmath,amssymb}
\usepackage{color}

\newcommand{\be}{\begin{equation}}
\newcommand{\ee}{\end{equation}}
\newcommand{\bea}{\begin{eqnarray}}
\newcommand{\eea}{\end{eqnarray}}
\newcommand{\bi}{\begin{itemize}}
\newcommand{\ei}{\end{itemize}}

\newcommand\hyn{\textit{\rm -}}

\begin{document}


\title{Charmonium-nucleon potential from lattice QCD}


\author{Taichi Kawanai}
\email{kawanai@nt.phys.s.u-tokyo.ac.jp}

\author{Shoichi Sasaki}
\email{ssasaki@phys.s.u-tokyo.ac.jp}
\affiliation{Department of Physics, The University of Tokyo, \\
Hongo 7-3-1, Tokyo 113-0033, Japan}

\preprint{TKYNT-10-14}

\date{\today}

\begin{abstract}
 We present results for the charmonium-nucleon potential
 $V_{c\bar{c}\hyn N}(r)$ from quenched lattice QCD, which is calculated from
 the equal-time Bethe-Salpeter amplitude through the effective
 Schr\"odinger equation. 
 Detailed information of the low-energy interaction between the nucleon and charmonia ($\eta_c$ and $J/\psi$)
 is indispensable for exploring the formation of charmonium bound to nuclei. 
 Our simulations are performed at a lattice cutoff of $1/a\approx 2.1$  GeV in a spatial volume of
 $(3\;\text{fm})^3$ with the nonperturbatively ${\cal O}(a)$ improved Wilson
 action for the light quarks and a relativistic heavy quark action for
 the charm quark. We have found that the potential $V_{c\bar{c} \hyn N}(r)$
 for either the $\eta_c$ and $J/\psi$ states
 is weakly attractive at short distances and exponentially screened at
 large distances. The spin averaged $J/\psi$-$N$ potential is
 slightly more attractive than that of the $\eta_c$-$N$ case.
\end{abstract}

\pacs{11.15.Ha, 
      12.38.-t  
      12.38.Gc  
}

\maketitle


 Heavy quarkonium states such as charmonium
 ($c\bar{c}$) states do not share the same quark flavor with the nucleon
 ($N$). This suggests that the heavy quarkonium-nucleon interaction is
 mainly induced by the genuine QCD effect of multi-gluon
 exchange~\cite{{Brodsky:1989jd},{Brodsky:1997gh},{Luke:1992tm}}.
 Therefore the $c\bar{c}\hyn N$ system is ideal to study the effect of
  multi-gluon exchange between hadrons. As an analog of the van der
 Waals force, the simple two-gluon exchange contribution gives a weakly attractive, 
 but long-ranged interaction
 between the heavy quarkonium state and the
 nucleon~\cite{{Appelquist:1978rt},{Feinberg:1979yw}}. 
 However, the validity of calculations based on a
 perturbative theory is questionable for QCD where 
 the strong interaction influences the long distance region.
 
 The $c\bar{c}$-$N$ scattering at low energy has been studied 
 from first principles of QCD. The $s$-wave $J/\psi$-$N$ scattering
 length is about 0.1 fm by using QCD sum
 rules~\cite{Hayashigaki:1998ey} and  $0.71\pm 0.48$ fm ($0.70\pm 0.66$
 fm for $\eta_c$-$N$) by lattice QCD~\cite{Yokokawa:2006td}, while it is
 estimated as large as 0.25 fm from the gluonic van der Waals
 interaction~\cite{Brodsky:1997gh}. All studies suggest that the
 $c\bar{c}$-$N$ interaction is weakly attractive. This indicates that the
  formation of charmonium bound to nuclei is enhanced.
 In 1991, Brodsky {\it et al.} had argued that the $c\bar{c}$-nucleus
 ($A$) bound system may be realized for the mass number $A\ge 3$ if the
 attraction between the charmonium and the nucleon is sufficiently
 strong~\cite{Brodsky:1989jd,Wasson:1991fb}. Therefore, precise
 information on the $c\bar{c}$-$N$ potential $V_{c\bar{c} \hyn N}(r)$ is
 indispensable for exploring nuclear-bound charmonium states like
 the $\eta_c$-${}^{3}{\rm He}$ or $J/\psi$-${}^{3}{\rm He}$ bound state in
 few body calculations~\cite{Belyaev:2006vn}.  
 
 We recall the  recent great success of the $N$-$N$ potential from lattice
 QCD~\cite{Ishii:2006ec}. In this new approach, the potential between
 hadrons can be calculated from the equal-time Bethe-Salpeter (BS)
 amplitude through an effective Schr\"odinger equation. Thus, direct
 measurement of the $c\bar{c}$-$N$ potential is now feasible by using
 lattice QCD. It should be important to give a firm theoretical
 prediction about the nuclear-bound charmonium, which is possibly
 investigated by experiments at J-PARC and FAIR/GSI.
 
 The method utilized here to calculate the hadron-hadron potential
 in lattice QCD is based on the same idea originally applied for 
 the $N$-$N$ potential~\cite{{Ishii:2006ec},{Aoki:2009ji}}. 
 We first calculate the equal-time BS amplitude of two local 
 operators (hadrons $h_1$ and $h_2$) separated by given spatial
 distances ($r=|{\bf r}|$) from the following four-point correlation
 function:
 %
 %
 \begin{eqnarray}
 &&G^{h_1\hyn h_2}({\bf r}, t; t_2, t_1)\nonumber \cr
&&=
 \sum_{\bf x}\sum_{{\bf x}^{\prime}, {\bf y}^{\prime}}
 \langle {h_1}({\bf x}, t){h_2}({\bf x}+{\bf r}, t)\left({h_1}
 ({\bf x}^{\prime}, t_2){h_2}({\bf y}^{\prime},t_1)    
 \right)^{\dagger}\rangle,
 \end{eqnarray}
 where ${\bf r}$ is the relative coordinate of two hadrons at sink position ($t$).
 Each hadron operator at source positions ($t_1$ and $t_2$) 
 is separately projected onto a zero-momentum state by a summation 
 over all spatial coordinates ${\bf x}^{\prime}$ and ${\bf y}^{\prime}$.
 To avoid the Fierz rearrangement of two-hadron operators, it is better
 to set $t_2\neq t_1$. 
 Without loss of generality, we choose $t_2=t_1+1=t_{\rm src}$ hereafter.
 Suppose that $|t-t_{\rm src}|\gg 1$
 is satisfied, the correlation function asymptotically behaves as
 \begin{equation}
 G^{h_1\hyn h_2}({\bf r}, t ; t_{\rm src})\propto
 \phi_{h_1\hyn h_2}({\bf r}) e^{-E_{h_1\hyn h_2}(t-t_{\rm src})},
 \end{equation}
 where the ${\bf r}$-dependent amplitude, which is defined as
 \begin{equation}
 \phi_{h_1\hyn h_2}({\bf r})=
 \sum_{{\bf x}}\langle 0| {h_1}
 ({\bf x}){h_2}({\bf x}+{\bf r})|h_1 h_2;E_{h_1\hyn h_2}\rangle
 \end{equation}
 with the total energy $E_{h_1\hyn h_2}$ for the ground state of the 
 two-particle $h_1\hyn h_2$ state,
 corresponds to a part of the equal-time BS amplitude and is called  the BS wave
 function~\cite{{Luscher:1990ux},{Aoki:2005uf}}.  
 After an appropriate projection with respect to discrete rotation
 \begin{equation}
 \phi^{A_{1}^{+}}_{h_1\hyn h_2}(r)=
 \frac{1}{24}\sum_{\mathcal{R}\in O_h} \phi_{h_1\hyn
 h_2}(\mathcal{R}^{-1}\mathbf{r}),
 \end{equation}
 where $\mathcal{R}$ represents 24 elements of the cubic group $O_h$,
 one can get the BS wave function projected in the $A^+_1$ 
 representation, which corresponds to the $s$-wave in continuum theory at low energy. 
 Once the BS wave functions $\phi^{A_1^+}_{h_1\hyn h_2}(r)$ 
 are calculated in lattice simulations, the hadron-hadron ``effective'' central potential with 
 the energy eigenvalue $E$ of the stationary Schr\"odinger equation, can be obtained by
 \begin{equation}
  V_{h_1\hyn h_2}(r)-E=\frac{1}{2\mu}
   \frac{\nabla^2 \phi^{A_1^+}_{h_1\hyn h_2}(r)}
   {\phi^{A_1^+}_{h_1\hyn h_2}(r)},
   \label{Eq.Pot}
 \end{equation}
 where {\small$\mu$} is the reduced mass of the $h_1$-$h_2$ system and
 {\small $\nabla^2$} is defined by the discrete Laplacian with
 nearest-neighbor points.
 Although the energy eigenvalue $E$ is supposed to be 
 the energy difference between the total energy of two hadrons ($E_{h_1\hyn h_2}$) and the sum of
 the rest mass of individual hadrons ($M_{h_1}+M_{h_2}$), we instead determine $E$ with the condition of
 $\lim_{r \rightarrow \infty}\{V_{h_1\hyn h_2}(r)-E\}=0$~\cite{Aoki:2005uf}.
 More details of this method can be found in
 Ref.~\cite{Aoki:2009ji}.

 Let us first consider the low energy $\eta_c \hyn N$
 interaction, which does not possess a spin-dependent part.
 We use the conventional interpolating operators,
 $h_1(x)=\epsilon_{abc}(u_a(x)C\gamma_5 d_b(x))u_c(x)$ for the nucleon
 and $h_2(y)=\bar{c}_a(y)\gamma_5 c_a(y)$ for the $\eta_c$ state, where
 $a$,\ $b$ and $c$ are  color
 indices, and $C=\gamma_4\gamma_2$ is the charge conjugation matrix. 
 We have performed quenched lattice QCD simulations on two different
 lattice sizes, $L^3\times T=32^3\times 48$ and $16^3\times 48$,
 with the single plaquette gauge action at $\beta=6/g^2=6.0$, which
 corresponds to a lattice cutoff of $a^{-1} \approx 2.1$ GeV according
 to the Sommer scale~\cite{Sommer:1993ce,Guagnelli:1998ud}.
 Our main results are obtained from the data taken on the larger lattice
 ($La\approx$ 3.0 fm). A supplementary data with a smaller lattice size
 ($La\approx 1.5$ fm) are used for a test of  finite-size effects. The
 number of statistics is ${\cal O}(600)$ for $L=32$ and ${\cal O}(200)$ for
 $L=16$, respectively. The simulation parameters and the number of sampled
 gauge configurations are summarized in Table~\ref{table1}.

 \begin{table}[!t]
  \caption{
  Simulation parameters in this study. 
  The Sommer parameter $r_0=0.5$
  fm is used to fix the scale~\cite{Sommer:1993ce,Guagnelli:1998ud}.
  \label{table1}
  }
  \begin{ruledtabular}
  \begin{tabular}{lccccc}
  \hline
   & $a$ & $a^{-1}$ & Lattice size & $\sim La$ & \cr
   $\beta$ & [fm] & [GeV] & ($L^3 \times T$) & [fm] & Statistics \cr
   \hline
   6.0 & 0.0931 & 2.12 & $16^3\times 48$ & 1.5 & 200 \cr
   &&&$32^3\times 48$ & 3.0 & 602\cr
   \hline
  \end{tabular}
  \end{ruledtabular}
 \end{table}

  We use non-perturbatively ${\cal O}(a)$ improved Wilson fermions 
  for light quarks ($q$) \cite{Luscher:1996ug} and a relativistic
  heavy quark (RHQ) action for the charm quark ($Q$)~\cite{Aoki:2001ra}.
  The RHQ action is a variant of the Fermilab approach
  \cite{ElKhadra:1996mp}, which can remove large discretization errors
  for heavy quarks. The hopping parameter is chosen to be
  $\kappa_q={0.1342,\ 0.1339,\ 0.1333}$, which correspond to
  $M_\pi={0.64,\ 0.72,\ 0.87}$ GeV ($M_N={1.43,\ 1.52,\ 1.70}$ GeV), 
  and $\kappa_Q=0.1019$ which is
  reserved for the charm-quark mass ($M_{\eta_c}=2.92$
  GeV and $M_{J/\psi}=3.00$ GeV)~\cite{Kayaba:2006cg}. 
  Each hadron mass is obtained by fitting the
  corresponding two-point correlation function with a single
  exponential form. We calculate quark propagators with wall sources,
  which are located at $t_{\text{src}}=5$ for the light quarks 
  and at $t_{\text{src}}=4$ for the charm quark, with  Coulomb gauge
  fixing. It is worth mentioning that Dirichlet boundary conditions 
  are imposed for quarks in the time direction in order to avoid 
  wrap-round effects, which are very cumbersome 
  in systems of more than two hadrons~\cite{Takahashi:2005uk}. 
  In addition, the ground state dominance in four-point functions is
  checked by an effective mass plot of total energies of the
  $c\bar{c}$-$N$ system.

  \begin{figure*}
   \centering
   \includegraphics[width=.49\textwidth]{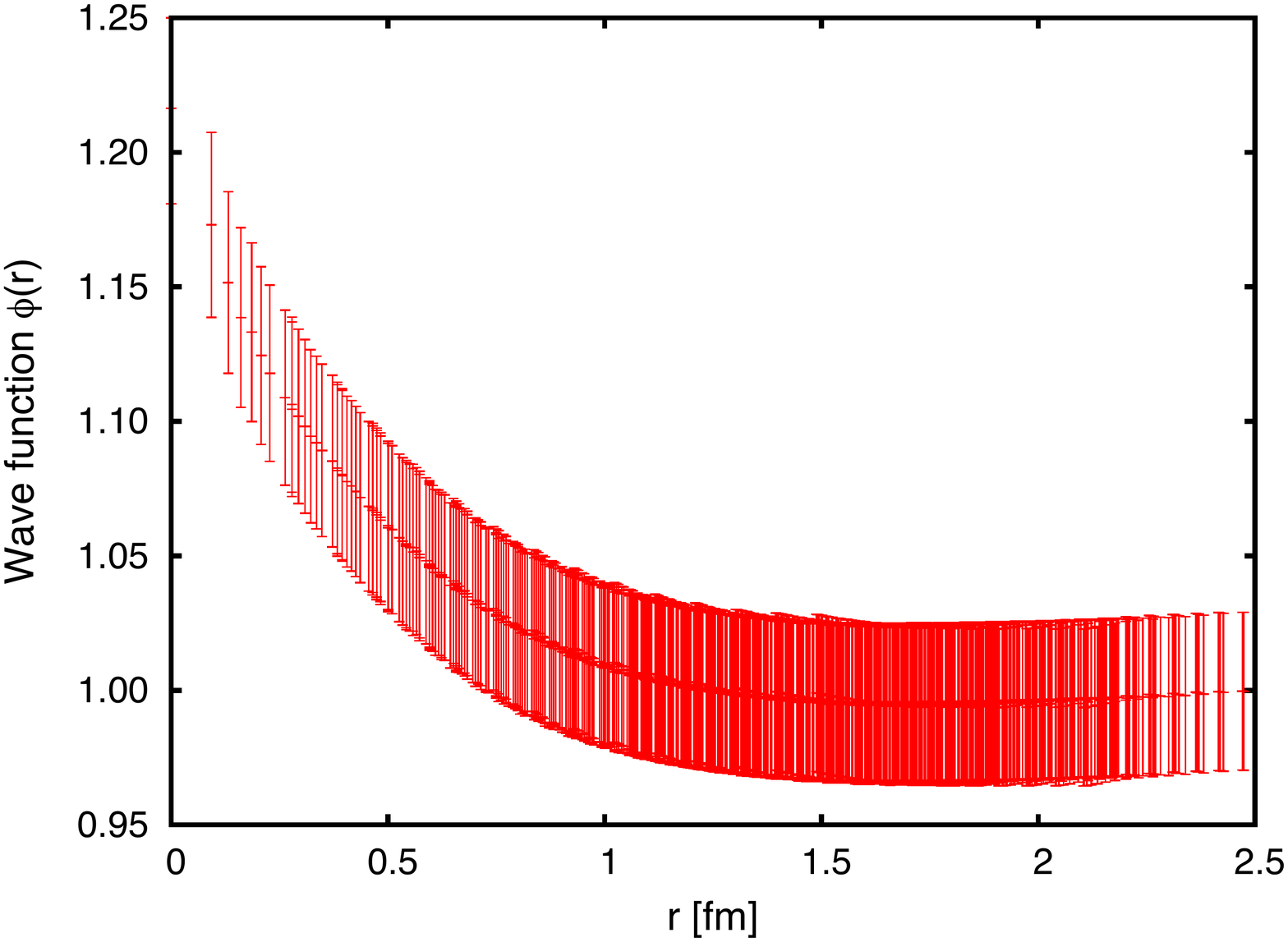}
   \includegraphics[width=.49\textwidth]{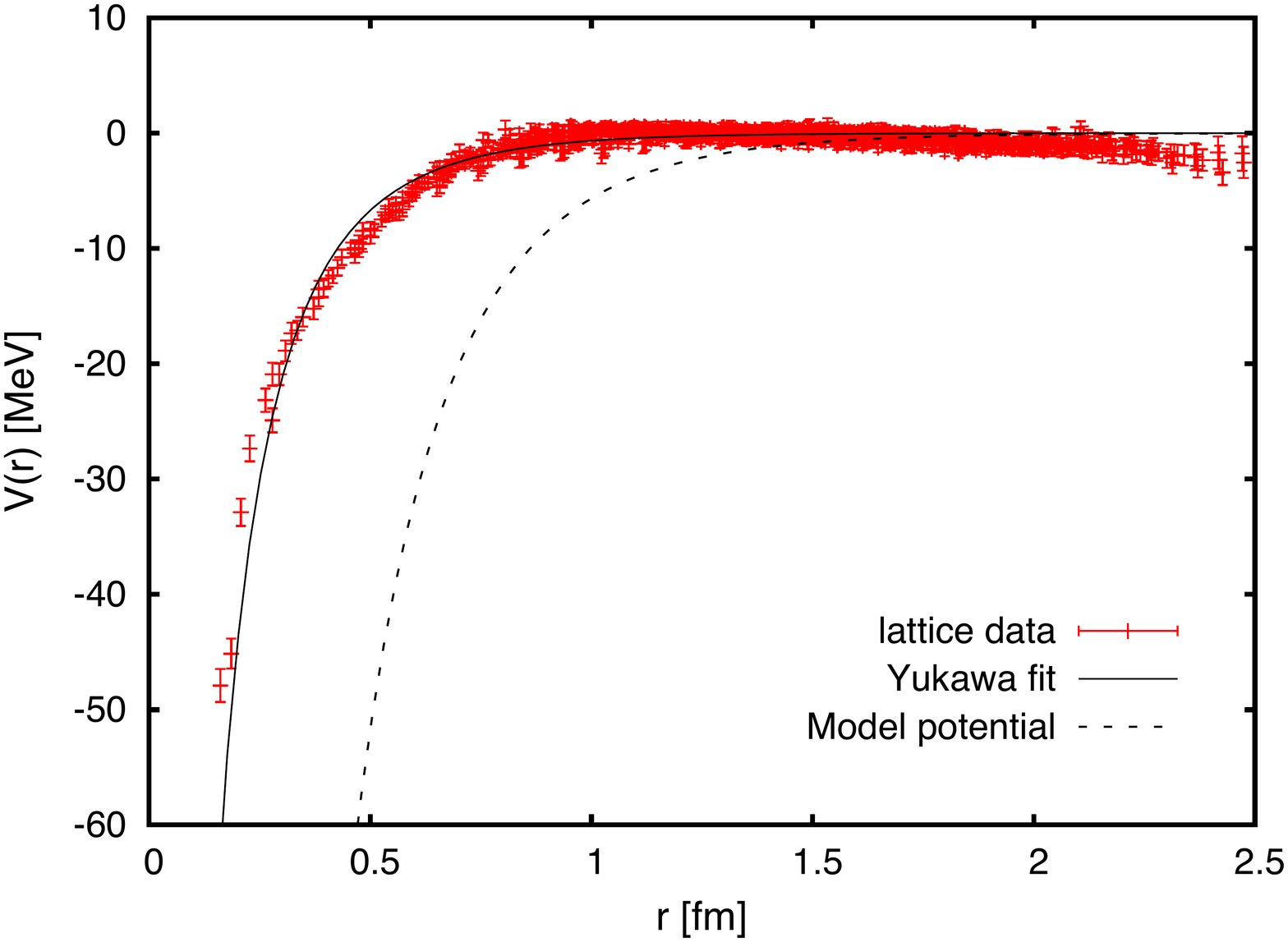}
   \caption{
   The wave function (left) and the effective central potential
   (right) in the $s$-wave $\eta_c$-$N$ system for $M_\pi= 0.64$
   GeV as a typical example.  In the right panel we fit a Yukawa
   potential (solid line) and compare with
   the phenomenological potential (dashed) adopted in
   Ref.~\cite{Brodsky:1989jd}.
   \label{Fig_results}}
  \end{figure*}
  The left panel of Fig.~\ref{Fig_results} shows a typical result of the
  projected BS wave function at the smallest quark mass, which is
  evaluated by a weighted average of data in the time-slice range of $16
  \leq t-t_{\text{src}}\leq 35$. The wave functions are normalized to unity
  at a reference point ${\bf r}=(16,16,16)$, which is supposed to be
  outside of the interaction region. As shown in Fig.~\ref{Fig_results},
  the wave function is enhanced from unity near the origin so that the
  low-energy $\eta_c\hyn N$ interaction is certainly attractive. This
  attractive interaction, however, is not strong enough to form a bound
  state as is evident from this figure, where the wave function is not
  localized, but extends to large distances.
  
  In the right panel of Fig.~\ref{Fig_results}, we show the effective
  central $\eta_c\hyn N$ potential, which is evaluated by the wave
  function through Eq.~(\ref{Eq.Pot}) with measured values of $E$ and $\mu$.
  As is expected, the $\eta_c\hyn N$ potential clearly exhibits an entirely
  attractive interaction between the charmonium and the nucleon 
  without any repulsion at either short or large distances. 
  The short range attraction is deemed to be a result of the absence of
  Pauli blocking,
  that is a relevant feature in this particular system of 
  the heavy quarkonium and the light hadron. The interaction is
  exponentially screened in the long distance region
  $r\gtrsim 1 \text{ fm}$~\footnote{
  The potential for $r\gtrsim 2 \text{ fm}$ deviates slightly from zero
  beyond statistical errors. This should be an artifact of the finite cubic 
  box of $(3.0 \text{ fm})^3$. 
  It is worth noting that the $r$-dependence of the potential in the region of 
  $r\gtrsim 1.5 \text{ fm}$ is mainly determined by off-axis separations.
  The difference between on and off-axis separations can be caused by
  a rotational symmetry breaking in the finite cubic box.
  }.
  This is consistent with the expected behavior of the color van der Waals
  force in QCD, where
  the strong confining nature of color electric fields must
  emerge~\cite{{Feinberg:1979yw},{Matsuyama:1978hf}}. The
  exponential-type damping in the color van der Waals force is hardly
  introduced by any perturbative arguments. 
      
  In detail, a long-range screening of the color van der Waals force is
  confirmed by the following analysis. We have tried to fit data with two
  types of fitting functions: i) exponential type functions
  $-\exp(-r^m)/r^n$, which include the Yukawa form ($m=1$ and $n=1$), and ii)
  inverse power law  functions $-1/r^n$, where $n$ and $m$ are not
  restricted to be integers. The former case can easily accommodate a
  good fit with a small $\chi^2$/ndf value, while in the latter case we
  cannot get any reasonable fit. For example, the functional forms $-\exp(-
  r)/r$ and  $-1/r^7$ give $\chi^2/\mbox{ndf}\simeq 2.5$ and $34.3$
  , respectively. It is clear that the long range force induced
  by a normal ``van der Waals'' type interaction based on two-gluon
  exchange~\cite{Feinberg:1979yw} is non-perturbatively screened.
  
  If we adopt the Yukawa form $-\gamma e^{-\alpha r}/r$ to fit our data
  of $V_{c\bar{c}\hyn N}(r)$, we obtain $\gamma\sim 0.1$ and $\alpha \sim 0.6$
  GeV. These values should be compared with the phenomenological
  $c\bar{c}$-$N$ potential adopted in Refs.~\cite{{Brodsky:1989jd}},
  where the parameters ($\gamma=0.6$, $\alpha=0.6$ GeV) are barely fixed by a 
  Pomeron exchange model. The strength of the Yukawa potential $\gamma$
  is six times smaller than the phenomenological value, while the Yukawa
  screening parameter $\alpha$ obtained from our data is
  comparable. The $c\bar{c}\hyn N$ potential obtained from
  lattice QCD is rather weak.

  \begin{figure*}
  \centering
  \includegraphics[width=.48\textwidth]{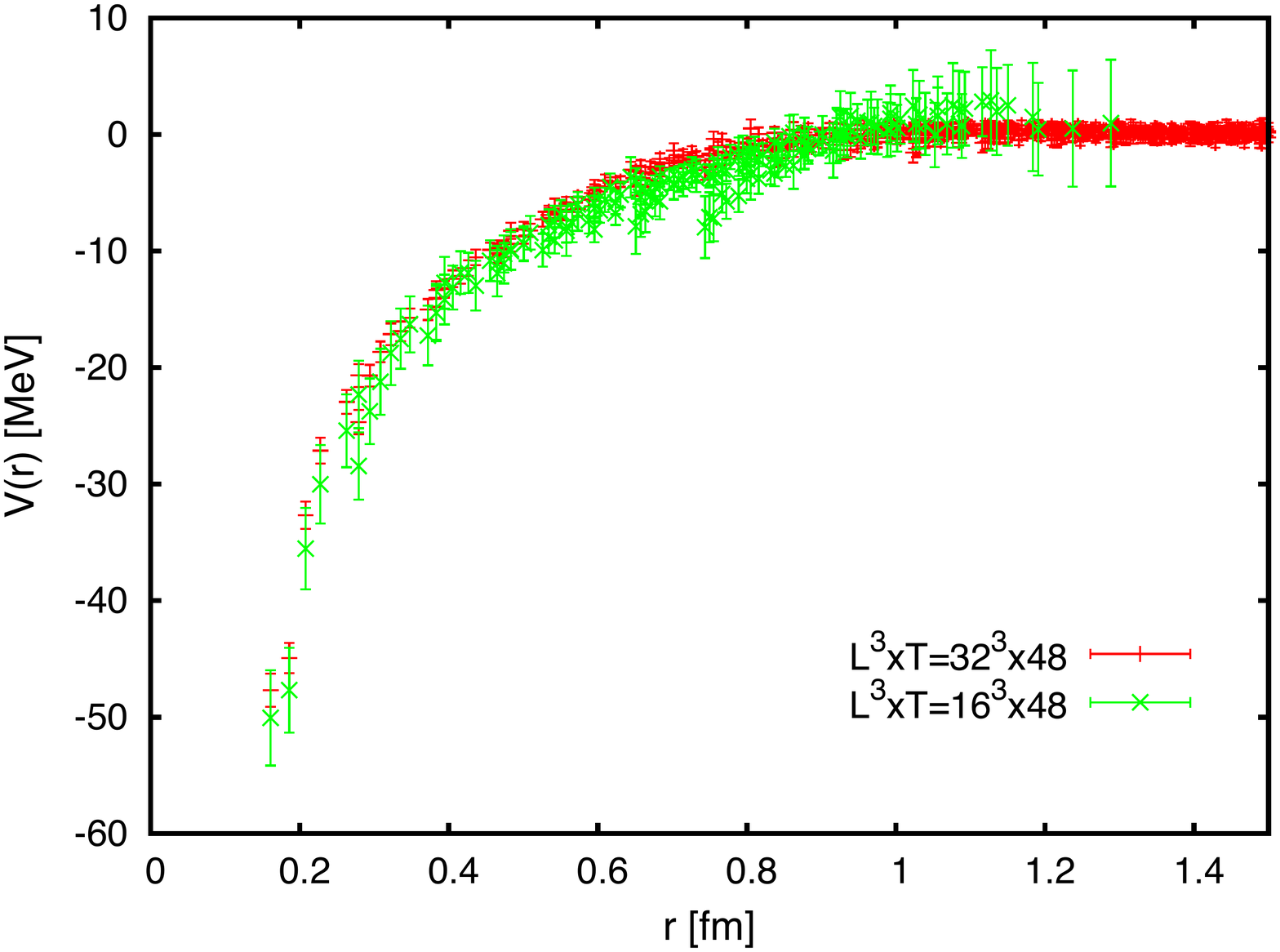} 
  \includegraphics[width=.48\textwidth]{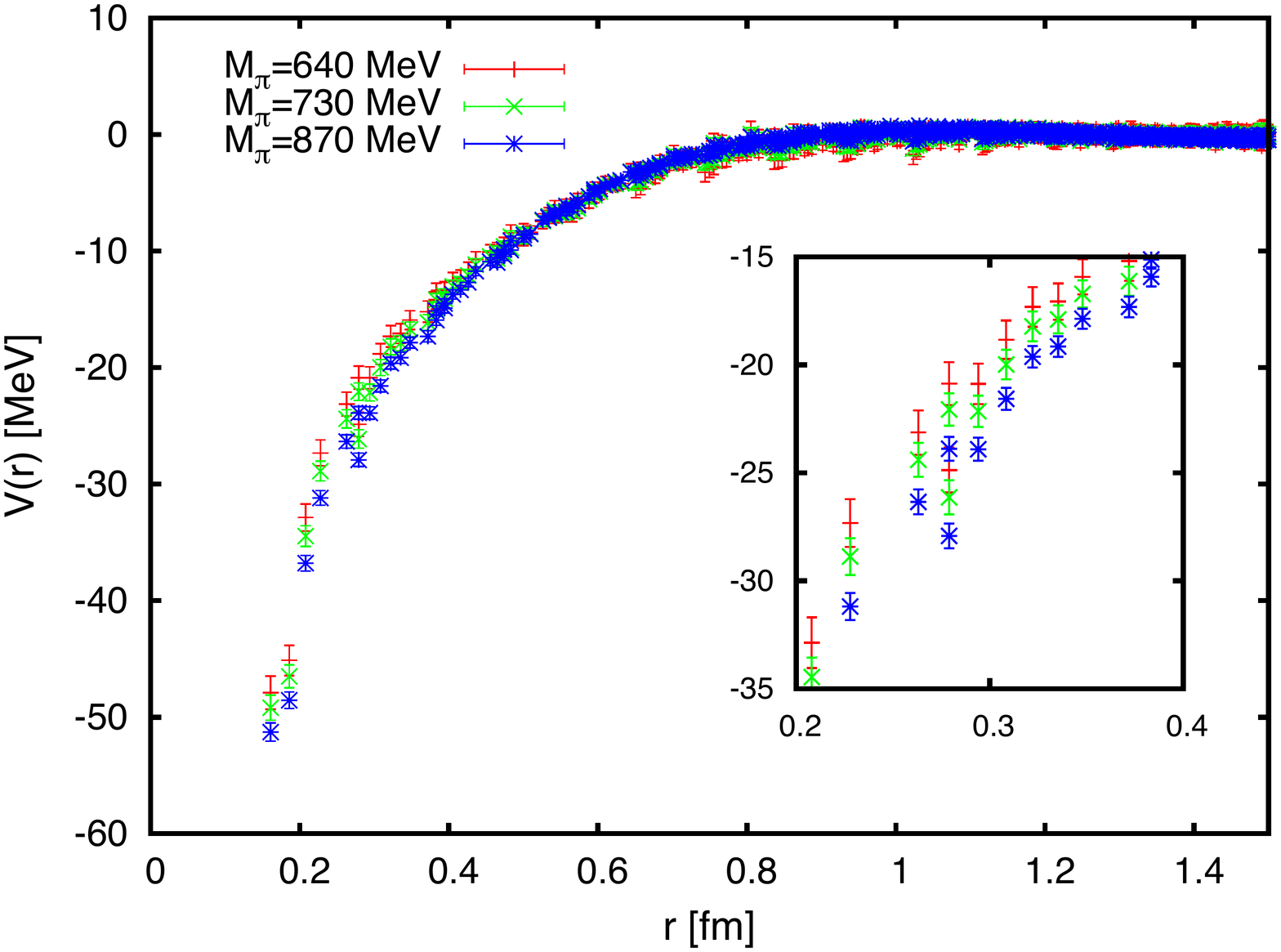}
  \caption{The volume dependence (left) and the quark-mass dependence
   (right) of the $\eta_c\hyn N$ potential.
  \label{dependence}}
  \end{figure*}
 We next show both finite-size and quark-mass dependence
 of the $\eta_c\hyn N$ potential in Fig.~\ref{dependence}. Firstly, as
 shown in the left panel of Fig.~\ref{dependence}, there is no
 significant difference between potentials computed from lattices with
 two different spatial sizes ($La\approx 3.0$ and 1.5 fm). This
 observation is simply due to the fact that the $\eta_c$-$N$
 potential is quickly screened to zero and turns out to be somehow short
 ranged. In principle, the short range part of the potential, which is
 represented by ultraviolet physics, should be insensitive to
 the spatial extent associated with an infrared cutoff. As a result, it is
 assured that the larger lattice size is large enough to study the
 $\eta_c$-$N$ system. 
 
 No large quark-mass dependence is also observed in the right panel of Fig.~\ref{dependence}. 
  This is a non-trivial feature since there is an explicit dependence on the
reduced mass $\mu$ in the definition of the effective central potential
 (\ref{Eq.Pot}).
 However, if one recalls that the $c\bar{c}\hyn N$ interaction is mainly governed by multi-gluon exchange,
 the resulting potential is expected to be less sensitive to the reduced
  mass of the considered system
 ignoring the internal structures of the $\eta_c$ and nucleon states.  

 If one takes a closer look at the inset of~Figure~\ref{dependence}, it
 is found that the nature of the attractive interaction 
 in the $\eta_c$-$N$ system tends to get slightly weaker as the light
 quark mass decreases. Does this
 mean that the strength of the $\eta_c$-$N$ potential at the physical
 point becomes much weaker than
 what we measured at the quark mass simulated in this study? The answer
 to this question is not simple.
 It is worth to remember that the ordinary van der Waals interaction is
 sensitive to the size of
 the charge distribution, which is associated to the dipole size. 
 Larger dipole size yields stronger interaction.  
 We may expect that the size of the nucleon becomes large as the light
 quark mass decreases.
 However, the very mild but opposite quark-mass dependence observed here
 does not 
 accommodate this expectation properly. 

 Recent detailed studies of nucleon form factors tell us that 
 the root mean-square (rms) radius of the nucleon, which is a typical
 size of the nucleon,
 shows rather mild quark-mass dependence and its value is much smaller than 
 the experimental value up to at $M_{\pi}\sim 0.3$ GeV (for example, see~\cite{Yamazaki:2009zq}). 
 At the chiral limit in baryon chiral perturbation theory
 the rms radius is expected to diverge logarithmically~\cite{Beg:1973sc}.
 This implies that the size of the nucleon increases drastically in
 the vicinity of
 the physical point. It may be phenomenologically regarded as the
 ``pion-cloud'' effect.
 We notice here that our simulations are performed in the quenched
 approximation and at rather 
 heavy quark masses. We speculate that the $c\bar{c}$-$N$ potential from
 dynamical
 simulations would become more strongly attractive in the vicinity of
 the physical point, where the size of the nucleon is much larger than
 at the quark mass simulated in this study.

  \begin{figure}
  \centering
  \includegraphics[width=.48\textwidth]{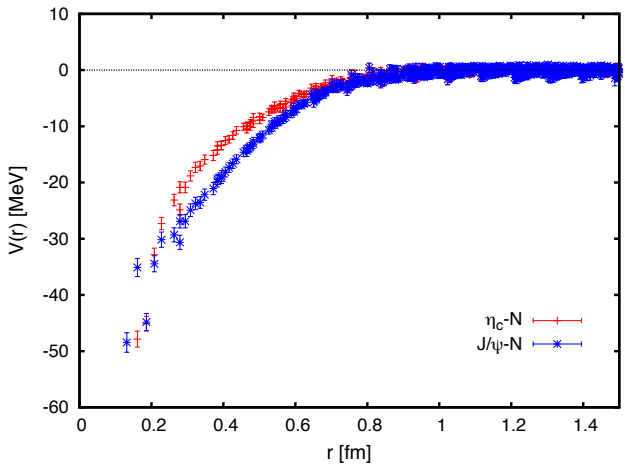} 
  \caption{The central and spin-independent part of the $J/\psi$-$N$
  potential at $M_\pi = 640$ MeV. The $\eta_c$-$N$ potential is
  also included for comparison.
  \label{jpsi}}
  \end{figure}

  We also have calculated the $J/\psi$-$N$ potential.
  It should be noted here that different total spin states 
  are allowed as spin-1/2 and spin-3/2 states in the $J/\psi$-$N$ system. 
  This fact introduces a little complexity regarding the spin-dependence on the $J/\psi$-$N$ potential.
  The interpolating operator of the $J/\psi$ state is defined 
  as $h_{2, i}(y)=\bar{c}_a(y)\gamma_i c_a(y)$, which carries the specific
  spin polarization direction.
  Therefore, the four-point correlation function for the $J/\psi$-$N$ state
  becomes a matrix form with spatial Lorentz indices, $G_{ij}^{h_1\hyn
  h_2}=\langle h_1 h_{2, i} (h_1 h_{2, j})^\dagger \rangle$. 
  It can be expressed by an orthogonal sum of spin-1/2 and spin-3/2
  components~\cite{Yokokawa:2006td}:
  $G_{ij}^{h_1\hyn
  h_2}=G^{1/2}\mathcal{P}^{1/2}_{ij}+G^{3/2}\mathcal{P}^{3/2}_{ij}$
  where proper spin projection operators for the spin-$1/2$ and spin-$3/2$
  contributions are given by $\mathcal{P}^{1/2}_{ij}=\frac{1}{3}\gamma_i\gamma_j$
  and $\mathcal{P}^{3/2}_{ij}=\delta_{ij}-\frac{1}{3}\gamma_i\gamma_j$
  in the center of mass frame~\cite{Benmerrouche:1989uc}.
  Then, each spin part can be projected out as 
  $G^{1/2}=\sum_{i,j}\mathcal{P}^{1/2}_{ij}G_{ji}^{h_1\hyn h_2}$
  and $G^{3/2}=\frac{1}{2}\sum_{i,j}\mathcal{P}^{3/2}_{ij}G_{ji}^{h_1\hyn
  h_2}$
  where the indices $i$ and $j$ are also summed over all spatial directions.
  
  As a result, we can obtain the BS wave function and the resulting $J/\psi$-$N$
  potential for each spin channel. Although the lower-spin state (spin-1/2) is not free
  from the contamination of the $\eta_c$-$N$ state through channel mixings~\cite{Yokokawa:2006td}, 
  we simply consider the spin averaged four-point correlation function for 
  the $J/\psi$-$N$ system as
  \begin{equation}
   G^{J/\psi \hyn N}_{\text{ave}}=\frac{1}{3}G^{1/2}+\frac{2}{3}G^{3/2}
    =\frac{1}{3}\sum_{i}G_{ii}^{J/\psi \hyn N}.
  \end{equation}
  This procedure may provide only a spin-independent part of the $J/\psi$-$N$ potential
  through the same analysis applied to the $\eta_c$-$N$ system. 

  We show our result of the $J/\psi$-$N$ potential in Fig.~\ref{jpsi} 
  where the $\eta_c$-$N$ potential is included for comparison.
  The $J/\psi$-$N$ potential shows short-range attraction. Similar to what
  we discussed in the $\eta_c$-$N$ case it does not have a normal ``van
  der Waals type'' $-1/r^n$ behavior.
  There is no qualitative difference between the $\eta_c$-$N$ and $J/\psi$-$N$
  potentials. Quantitatively, the attractive interaction observed in the
  $J/\psi$-$N$ potential is rather
  stronger than that of the $\eta_c$-$N$ system as shown in
  Fig.~\ref{jpsi}, though
  it is still not strong enough to form a bound state in the
  $J/\psi$-$N$ system.

  What is a possible origin of the stronger attraction appearing in the
  $J/\psi$-$N$ system? As was discussed previously, the attractive
  interaction in the $c\bar{c}$-$N$ system tends to be slightly stronger
  as the light quark mass increases. It should be recalled that the
  reduced mass of the $c\bar{c}$-$N$ system is also changed through a
  variation of the light quark mass. Supposed that the strength of the
  $c\bar{c}$-$N$ potential depends simply on the reduced mass of the
  $c\bar{c}$-$N$ system, a mass difference of the $\eta_c$ and $J/\psi$
  state may account for the difference between the $\eta_c$-$N$ and
  $J/\psi$-$N$ potentials. However, this is not the case. As shown in
  the inset of Figure~\ref{dependence}, the $\eta_c$-$N$ potential gets
  deeper, when the reduced mass of the $\eta_c$-$N$ system is increased
  by about 10\% through a variation of the light quark mass. On the
  other hand, although the reduced mass receives only about 1\% gains in
  the $J/\psi$-$N$ system relative to the $\eta_c$-$N$ system, the
  difference between the $\eta_c$-$N$ and $J/\psi$-$N$ potentials shown
  in Fig.~\ref{jpsi} is much bigger than the reduced mass dependence
  observed in Fig.~\ref{dependence}. This may indicate that the stronger
  interaction in the $J/\psi$-$N$ system than in the $\eta_c$-$N$ system 
  is caused by some dynamics associated with the structure of quarkonia. 
   
  We again recall that the ordinary van der Waals interaction is sensitive to 
  the size of the charge distribution. Therefore, what we observed here
  is intuitively 
  accounted for by the simple speculation that the size of the $J/\psi$
  state is larger than 
  that of the $\eta_c$ state.  Nevertheless, similar to the $\eta_c$-$N$
  system in the $J/\psi$-$N$ no
  appreciable finite-size effects are observed.
    
 We have studied the charmonium-nucleon potential
 $V_{c\bar{c}\hyn N}(r)$ in quenched lattice QCD.
 Potentials between charmonia ($\eta_c$ and $J/\psi$) and the nucleon 
 are calculated from the equal-time BS amplitude 
 through the effective Schr\"odinger equation.  
 We have found that the central and spin-independent potential
 $V_{c\bar{c}\hyn N}(r)$ 
 in both the $\eta_c$-$N$ and $J/\psi$-$N$ systems is weakly attractive at
 short distances and exponentially screened at large distances. 
 It is observed that both potentials have no appreciable finite-size 
 dependence and no significantly large quark-mass dependence within the
 pion mass region $640\textrm{MeV}\leq M_\pi\leq 870\textrm{MeV}$.
  The $J/\psi$-$N$ system is slightly stronger attractive than the
  $\eta_c$-$N$ system.
  This should be accounted for by the dynamics associated
 with the structure of quarkonia. 
 
 In order to make a reliable prediction about the nuclear-bound charmonium,
 an important step in the future is clearly an extension to dynamical
 lattice QCD simulations in the lighter quark mass region. Such planning is
 now underway. Once we obtain a realistic potential, we will proceed
 exploring the nuclear-bound charmonium state with theoretical inputs of the
 charmonium-nucleon potential by using the exact few-body calculation.
 
 We would like to thank T. Hatsuda for helpful suggestions and fruitful discussions.
 We also thank to T. Misumi for his careful reading of the manuscript.
 T.K. is supported by Grant-in-Aid for the Japan Society for Promotion
 of Science (JSPS) Research Fellows (No.~22-7653).
 S.S. is  supported by the JSPS Grant-in-Aids for Scientific Research (C)
 (No.~19540265) and Scientific Research on Innovative Areas
 (No.~21105504). 
 Numerical calculations reported here were carried out
 on the PACS-CS supercomputer at CCS, University of Tsukuba and also on
 the T2K supercomputer at ITC, University of Tokyo.

\end{document}